\documentclass[preprint,prb,aps,amsmath,amssymb,color,superscriptaddress]{revtex4}
\usepackage{stmaryrd}
\usepackage{amsmath}
\usepackage{amssymb}
\usepackage{graphicx}
\usepackage{dcolumn}
\usepackage{bm}
\usepackage{multirow}
\usepackage{tabularx}
\usepackage{epsfig}

\begin{document}
\begin{titlepage}
\title{Excitonic Instability and Electronic Property of Two-dimensional AlSb Limit}
\author{Shan Dong}
\affiliation{Key Lab of advanced optoelectronic quantum architecture and measurement (MOE), and Advanced Research Institute of Multidisciplinary Science, Beijing Institute of Technology, Beijing 100081, China}
\author{Yuanchang Li}
\email{yuancli@bit.edu.cn}
\affiliation{Key Lab of advanced optoelectronic quantum architecture and measurement (MOE), and Advanced Research Institute of Multidisciplinary Science, Beijing Institute of Technology, Beijing 100081, China}
\date{\today}

\begin{abstract}
Motivated by the recent synthesis of two-dimensional monolayer AlSb, we theoretically investigate its ground state and electronic properties using the first-principles calculations coupled with Bethe-Salpeter equation. An excitonic instability is revealed as a result of larger exciton binding energy than the corresponding one-electron energy gap by $\sim$0.1 eV, which is an indicative of a many-body ground state accompanied by spontaneous exciton generation. Including the spin-orbit coupling is proven to be a must to correctly predict the ground state. At room temperature, the two-dimensional monolayer AlSb is a direct gap semiconductor with phonon-limited electron and hole mobilities both around 1700 cm$^2$/V$\cdot$s. These results show that monolayer AlSb may provide a promising platform for realization of the excitonic insulator and for applications in the next-generation electronic devices.
\end{abstract}

\maketitle
\draft
\vspace{2mm}
\end{titlepage}

\vspace{0.3cm}
\textbf{I. Introduction}
\vspace{0.3cm}

The successful isolation of graphene in 2004 demonstrated the possibility to synthesize atomically thin materials from the layered van der Waals solids by exfoliation. Since then, there is growing theoretical and experimental interest in the research of other two-dimensional materials such as hexagonal-BN, transition metal dichalcogenides, phosphorene, metal halides, \emph{etc}. These crystals exhibit many fascinating electronic, magnetic and optic properties that are absent in their bulk counterparts, and thus hold a great application potential ranging from field-effect transistors to spin- and valley-tronics to photocatalytic water splitting\cite{Butler,Bernardi,WangG}. They also provide new opportunities for some long-standing physical problems, for example, the excitonic insulator.

Excitonic insulator was proposed by theoretical physicists more than 50 years ago\cite{Mott,Knox,Kohn,Halperin}. It is originated from the excitonic instability caused by the larger exciton binding energy than the corresponding one-electron energy gap, alongside with spontaneous exciton condensation. It has a many-body ground state similar to a superconductor and spontaneous symmetry-breaking occurs\cite{KohnRMP}. However, to date, the excitonic insulator has remained a mystery and there is still a lack of compelling evidence in experiment\cite{Kogar}. Atomically thin materials usually have a significantly enhanced exciton binding energy owing to their weak electron-hole screening interaction as compared to their bulk counterparts\cite{WangG,usPRL}. They naturally provide more opportunities for the occurrence of excitonic instability and the realization of excitonic insulator state\cite{usEI,usHEI,usgraphone,usDong,Varsano,LiZ}. In traditional bulk materials, the excitonic insulator is often limited to narrow-gap semiconductors or semi-metals, while in the two-dimensional limit, the existence in moderate or wide gap semiconductors is also discussed by recent works\cite{usEI,usHEI,usgraphone}.

Another outstanding advantage of two-dimensional materials is their suppressed short-channel effects at the scaling limit\cite{Yan}, which makes them promising in building the next generation of field-effect transistors. Although graphene offers the ultra-high carrier mobility, the lack of an energy gap leads to a low on-off current ratio, limiting its device applications\cite{Liao}. Single-layer MoS$_2$, which is the mostly studied member of the transition metal dichalcogenides, possesses a gap $\sim$1.8 eV and shows $n$-type conductivity with a mobility $\sim$200 cm$^2$/V$\cdot$s \cite{Yoon,Kaasbjerg,Cai}. Monolayer phosphorene has a direct gap of 1.5 eV and shows high hole mobility\cite{Liu,Qiao}. The search for new two-dimensional semiconductors with suitable band gap and high mobility has been a never-ending pursue. Binary III-V semiconductors are one of the most important class of semiconducting materials with enormous technological applications. The bulk of these compounds generally crystalize in zinc-blende or wurtzite structure. Such a non-layered nature thus hinders to directly exfoliate their two-dimensional counterparts, which, instead, can be obtained via surface growth\cite{Butler,Balushi}. The study of two-dimensional form of III-V semiconductors is still in its infancy. Experimentally, Balushi et al.\cite{Balushi} synthesized a bilayer GaN on the SiC substrate via a migration-enhanced encapsulated growth technique. Theoretically, Lucking et al.\cite{Lucking} predicted that the III-V semiconductors stabilize in a double layer honeycomb structure at the ultrathin limit. Very recently, Qin et al.\cite{Qin} reported the successful growth of two-dimensional AlSb in the double-layer honeycomb structure through molecular beam epitaxy on graphene-covered SiC(0001), and a fundamental gap of 0.93 eV, similar to the Si, is found.

In this paper, we investigate the ground-sate geometric and electronic structure of monolayer AlSb using the first-principles calculations plus the Bethe-Salpeter equation (BSE). An excitonic instability is revealed because the ground-state exciton has a binding energy larger than the corresponding one-electron gap by $\sim$0.1 eV. This implies that the monolayer AlSb could be an intrinsic excitonic insulator. Our calculations indicate that spin-orbit coupling (SOC) must be involved in order to correctly predict the ground state of monolayer AlSb. Since the many-body ground state tends to become unstable above the phase transition temperature, we further calculate the intrinsic carrier mobility of monolayer AlSb within deformation potential theory. Both electron and hole mobilities are found to be on the level of $\sim$1700 cm$^2$/V$\cdot$s at room temperature, and hence monolayer AlSb may also serve as a promising candidate for several electronic device applications, such as a high-energy photon detector\cite{Yee}.

\vspace{0.3cm}
\textbf{II. Methodology and models}
\vspace{0.3cm}

First-principles calculations were carried out by using the QUANTUM ESPRESSO package\cite{QE} with the Perdew-Burke-Ernzerhof (PBE) \cite{PBE} exchange-correlation functional. A cut-off energy of 50 Ry was employed for optimized norm-conserving Vanderbilt pseudopotentials\cite{Hamann}. An $18\times 18\times 1$ $k$-point grid was used to sample the Brillouin zone. A vacuum space of 20 \AA\ along the direction normal to the double-layer plane was set to avoid spurious interactions between adjacent layers. In order to cure the band gap problem of the Kohn-Sham DFT, single-shot $G_0W_0$ calculations were performed by the YAMBO code\cite{yambo}. Dielectric functions and low-energy excitonic properties were calculated by solving the BSE with the Coulomb cutoff technique. Top four valence bands and bottom two conduction bands are chosen to build the BSE Hamiltonian. The same $k$ grid, 312 bands, and 12 Ry cutoff were used for the dielectric function matrix.

%fig01
\begin{figure}[tbp]
	\includegraphics[width=0.8\columnwidth]{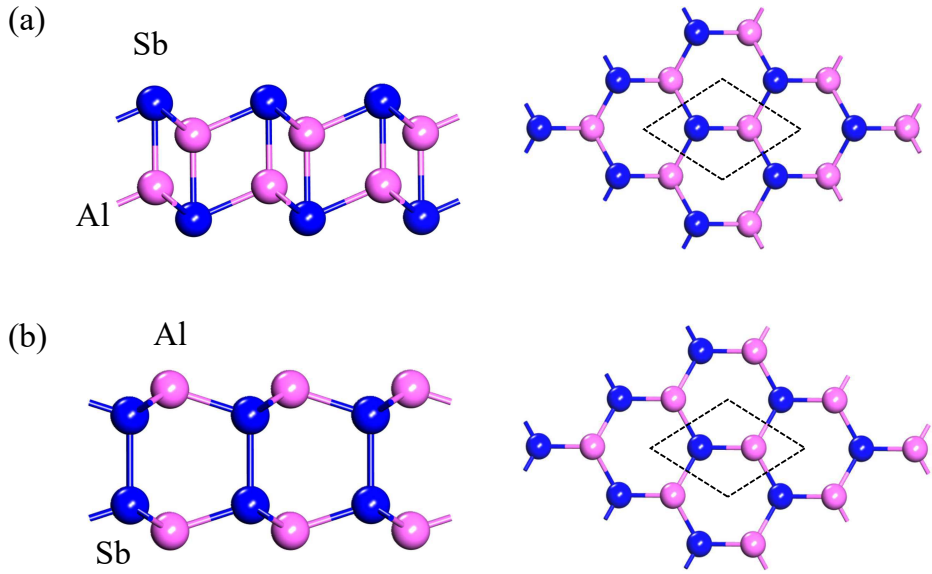}
	\caption{\label{fig:fig1} (Color online) Side views (Left panel) and top views (Right panel) of two kinds of double-layer configuration for two-dimensional AlSb limit. (a) Proposed by Lucking et al. in Ref. \onlinecite{Lucking}. (b) InSe-like structure with bonding between the same Sb atoms. Black dashed rectangles denote the unit cell.}
\end{figure}

\vspace{0.3cm}
\textbf{III. Results and discussion}
\vspace{0.3cm}

The essence that $p$-electrons tend to form covalent bonds often leads to a buckled geometry at the single-layer honeycomb limit when the compounds do not contain one of the first row elements\cite{Sahin}. For III-V semiconductors, this will be accompanied by a dipole moment along the direction perpendicular to the basal plane, which is not conductive to lower the system energy. In this sense, Lucking et al.\cite{Lucking} proposed  a double-layer honeycomb structure as illustrated in Fig. 1(a), to compensate the vertical dipole moments through an ``inter-layer" coupling, which consequently becomes more favorable in energy than the single-layer case. In fact, the double-layer honeycomb structure can also be presented in another form as depicted in Fig. 1(b). The latter is often taken by monolayer III-VI semiconductors like InSe\cite{Lei}, which differs from the one shown in Fig. 1(a) in the bonding between atoms of the same element. We compare the energetics of two-dimensional AlSb with the above two configurations. Our calculated results show that the InSe-like AlSb structures is energetically unfavorable by about 1.32 eV than the one proposed by Lucking et al.

%fig02
\begin{figure}[tbp]
	\includegraphics[width=0.8\columnwidth]{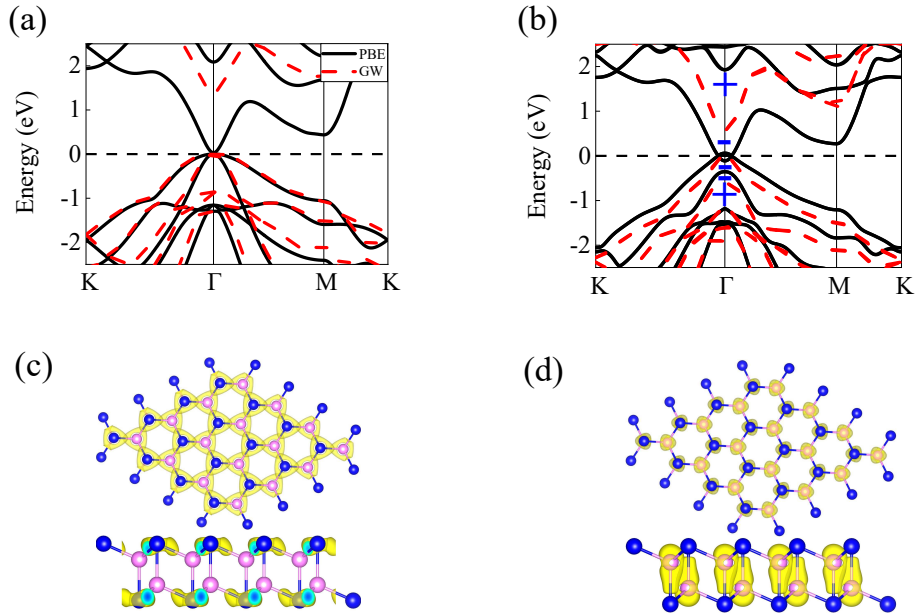}
	\caption{\label{fig:fig2} (Color online) (a)[(b)] One-electron band structure of monolayer AlSb without[with] the spin-orbit coupling. Black solid lines are from the PBE while red dashed lines are from the $G_0W_0$ on top of the PBE. The Fermi level is set to energy zero. Band parities are also calculated for the states near the Fermi level, and in (b) ``+" and ``-" denote the even and odd parity, respectively. (c)[(d)] Decomposed charge density for the top valence[bottom conduction] band at the $\Gamma$ point corresponding to the PBE band shown in (a), with an isosurface of 0.07 e/\AA$^3$.}
\end{figure}

Next, we explore the ground-state electronic properties of ultrathin AlSb with the configuration of Fig. 1(a). Shown in Fig. 2(a) is its band structure without considering the SOC. At the PBE level, it is a semiconductor with a direct gap very close to zero ($\sim$0.04 eV) at the $\Gamma$ point. This is in sharp contrast to the bulk AlSb which has an indirect-gap of 1.29 eV from the PBE\cite{Lucero} and 1.686 eV from the experiment\cite{Alibert}. The fact that the calculated gap of bulk AlSb  is $\sim$30\% smaller than the experimental value once again proves the persistent problem of PBE's gap underestimation. A similar trend is certainly anticipated for the two-dimensional limit, and the underestimation may be even more severe as a result of the reduced dimension. To this end, we carried out many-body $G_0W_0$ calculation and the corresponding band is plotted in Fig. 2(a) [Red dashed lines] for comparison. It gives rise to a quasi-particle gap of 1.35 eV, which is much larger than $\sim$0.04 eV by the PBE. Except for the gap size, the $G_0W_0$ band highly resembles the PBE one, including the band dispersion and the direct-gap characteristic.

We note, however, that the $G_0W_0$ gap of 1.35 eV becomes much larger than the experimental measurement of 0.93 eV\cite{Qin}. One possible reason is the neglect of SOC given that the compound contains heavy element Sb. Indeed, gap reduction resulted from the SOC has been reported in the bulk AlSb\cite{Ali}. As can be seen in Fig. 2(b), the band changes a lot when switching on the SOC. At the PBE level, it exhibits a metallic behavior and there is a small overlap between the top valence band and bottom conduction band, yielding a negative gap of -0.17 eV at the $\Gamma$ point. The $G_0W_0$ opens up a positive gap but the size of 0.74 eV is smaller by 0.19 eV than the experiment\cite{Qin}. Apart from the energy gap, another notable feature caused by the SOC is the lift of degeneracy at the $\Gamma$ point. This is especially significant for the top valence band, up to 0.7 eV (0.42 eV) by the $G_0W_0$ (PBE).

One can understand the SOC induced gap reduction as follows. Plotted in Figs. 2(c) and 2(d) are the decomposed charge densities for the band edge states from the PBE [see Fig. 2(a)]. One can see that the valence band maximum is dominantly comprised of the Sb orbitals [see Fig. 2(c)]. The Sb's $p_x$ and $p_y$ orbitals form a $\pi$-like bond and exhibit an in-plane distribution. On the other hand, the conduction band minimum is dominantly comprised of the Al orbitals [see Fig. 2(d)]. The Al's $s$ orbitals form a head-to-head $\sigma$-like bond and exhibit an out-of-plane distribution. Because the Sb has a much stronger SOC than that of the Al, the SOC induced splitting remarkably pushes up the top valence band, resulting in the narrowing of energy gap.

It is noted that for III-V semiconductors, the energy gap under two-dimensional limit is generally smaller than their bulk counterparts, for here observed AlSb or previously reported GaAs and others\cite{usEI,Lucking}. Nevertheless, the trend is reversed for II-VI and I-VII semiconductors\cite{Lucking}, as well as van der Waals layered crystals\cite{Qiao,Zhang,Gonzalez,Zhao}, that is, the energy gap usually increases from the bulk down to the monolayer as a result of the confinement effect. Such a difference is probably attributed to their unique double-layer honeycomb structure and deserves further exploration.

%fig03
\begin{figure}[tbp]
	\includegraphics[width=1\columnwidth]{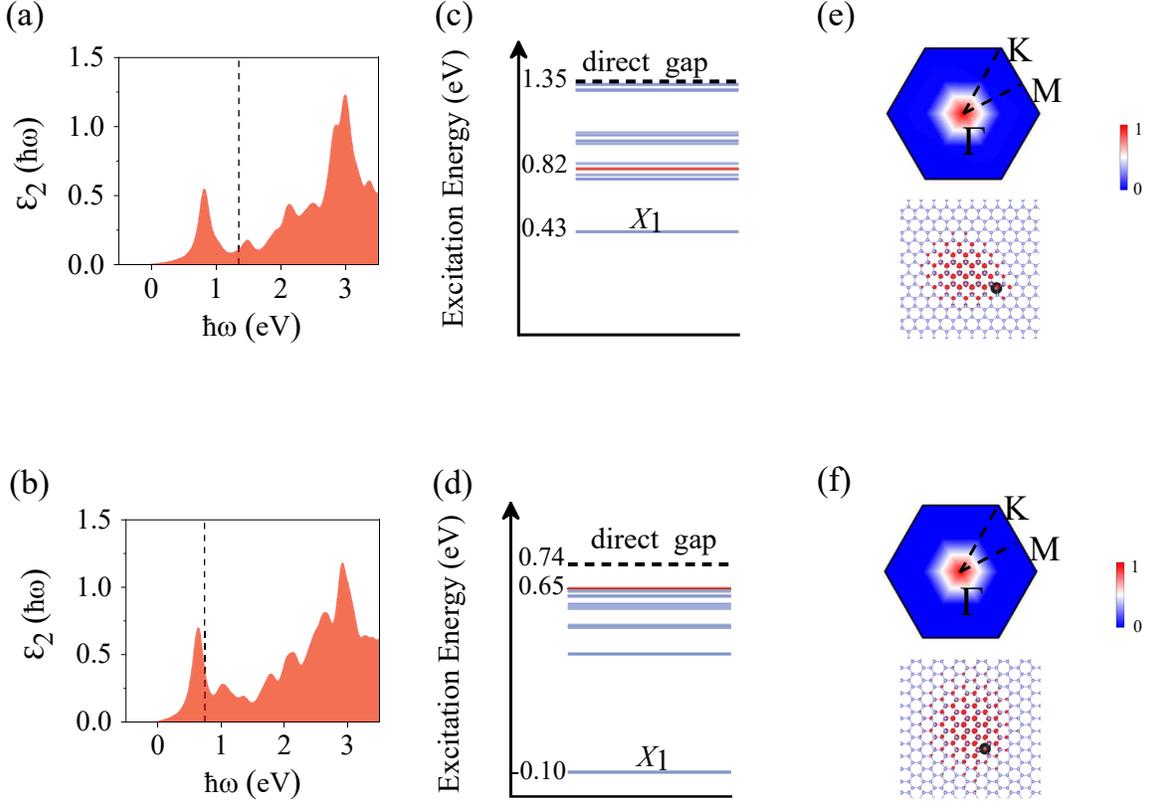}
	\caption{\label{fig:fig3} (Color online) (a)[(b)] Imaginary part ($\varepsilon_2$) of the BSE dielectric function, and (c)[(d)] all exciton states below the quasi-particle gap without[with] the spin-orbit coupling in the low-energy region. In (c) and (d), blue and red horizontal lines represent dark and bright excitons, and $X_1$ denotes the ground-state exciton with the lowest excitation energy. A negative excitation energy means spontaneous generation of the corresponding exciton. (e)[(f)] Reciprocal and real-space wave functions of the $X_1$ exciton without[with] the spin-orbit coupling. The density has been normalized by choosing the maximum value to be unity for reciprocal wave functions. While for the real-space plots, the isosurface corresponds to an electron density of 0.68 e/\AA$^3$. Black dots denote the hole positions.}
\end{figure}

The III-VI semiconductors have traditionally been used for luminescence. In combination with the direct-gap nature of monolayer AlSb, we turn to investigate its optical properties. In Figs. 3(a) and 3(b), we compare the optical adsorption spectra without and with the SOC, respectively. Both are calculated on top of the corresponding $G_0W_0$ bands. Not considering the SOC, the first absorption peak appears at 0.82 eV, which is located inside the quasi-particle gap. Hence, it corresponds to an exciton adsorption, with a binding energy of 0.53 eV. When the SOC is included, the first adsorption peak is located at 0.65 eV, which is just 0.09 eV lower than the quasi-particle gap.

Despite the lack of any spectral signature, the dark exciton states are also attractive in low-dimensional systems, especially for collective effects such as a Bose-Einstein condensation\cite{usEI,usHEI,usgraphone,usDong,Beian}. In this regard, we present all low-energy excitons of monolayer AlSb below the quasi-particle gap in Figs. 3(c) and 3(d), no matter optically active or inactive. Switching off the SOC, as illustrated in Fig. 3(c), five dark exciton states emerge below the first bright exciton. The minimum energy required to generate an exciton (dark) is 0.43 eV, which is 0.39 eV smaller than the required by photoexcitation. When the SOC is taken into account, many dark exciton states emerge below the first bright exciton [see Fig. 3(d)]. But in this case, to our surprise, the lowest exciton state has a negative excitation energy, namely, -0.10 eV. This means that spontaneous exciton production will lower the system energy. As a result, an excitonic instability would occur relative to the one-electron band structure, pointing towards the formation of a many-body ground state, i.e., the excitonic insulator\cite{Kohn}.

Two points are noteworthy for here predicted excitonic instability. On the one hand, it is the role played by the SOC. Above results clearly show the necessity of including SOC. Otherwise, ones would wrongly predict an one-electron ground state. In order for a better understanding, we compare the SOC impact on properties of the ground-state exciton, denoted as the $X_1$ in Figs. 3(c) and 3(d). First, it is doubly degenerate under both cases, in connection with the transition between valence band maximum and conduction band minimum. Second, it is not the spin selection rule that forbids the dipole transition as the $X_1$ remains to be dark in the absence of the SOC. Previous study\cite{usEI} in similar monolayer GaAs indicated that the elimination comes from the band parity. For this reason, we calculate the parity for relevant states. It turns out that the frontier states all possess the same odd parity as shown in Fig. 2(b), so the transitions between them are dipole forbidden, producing these low-energy dark excitons. Third, we compare the reciprocal and real-space wave functions of the $X_1$ exciton in Figs. 3(e) and 3(f). No apparent difference is found. Both are concentrated on the $\Gamma$ point in the Brillouin zone and display an isotropic $s$-wave character. Localization in the reciprocal space implies delocalization in the real space. Indeed, they become highly extended, over more than ten unit cells. Relatively speaking, the exciton wave function is more delocalized when the SOC is considered. This is consistent with the intuition that the exciton binding energy of 0.84 eV is slightly less than that of 0.92 eV given by ignoring the SOC.

On the other hand, we have to keep in mind that the $G_0W_0$ plus SOC calculations lead to an underestimated one-electron gap for monolayer AlSb, namely, 0.74 eV vs. 0.93 eV of experiment. Needless to say, this will profoundly affect the system screening interaction, and may even reverse the relative size of one-electron energy gap and exciton binding energy, changing the conclusion of excitonic instability. Although the $G_0W_0$ calculation on top of the HSE input succeeded to produce a one-electron gap comparable to the experiment\cite{Qin}, directly solving the BSE on this level is, however, unfortunately, too computationally demanding to carry out for us at present. Herein we circumvent the heavy-cost calculations and chose an alternative way of using the scissor operator, which is introduced both for the response function and diagonal part of the BSE kernel\cite{Sangalli}.

%fig04
\begin{figure}[tbp]
    \includegraphics[width=0.7\columnwidth]{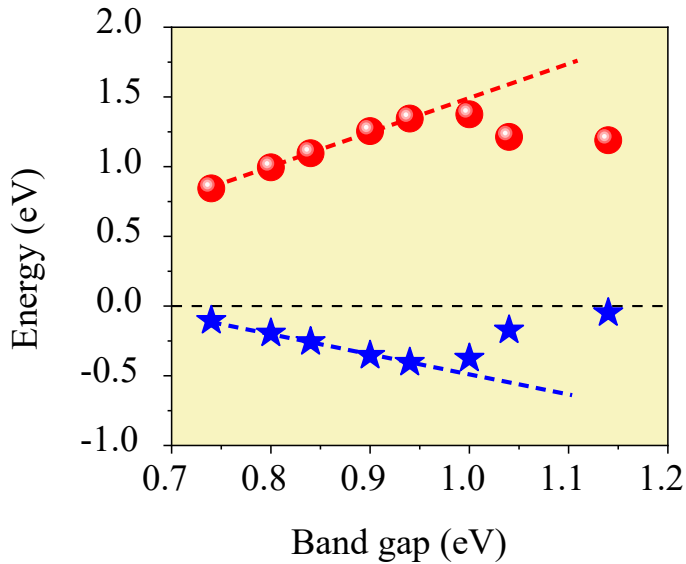}
    \caption{\label{fig:fig4} (Color online) Dependence of excitation (Red balls) and binding (Blue stars) energy for the $X_1$ exciton as a function of the quasi-particle gap that is corrected by a scissor operator. Red and blue dashed lines are a guide for the eye. Black horizontal dashed line indicates energy zero.}
\end{figure}

In Fig. 4, we summarize the dependence of exciton excitation/binding energy as a function of the quasi-particle gap (after applying a scissor correction). Above all, the exciton excitation energy is always negative in the whole gap range studied, from 0.74 eV to 1.14 eV. Note that the gap of 1.14 eV has exceeded the experimental value of 0.93 eV by 0.21 eV but the exciton excitation energy remains to be negative, i.e., -0.05 eV. In other words, although the calculation error on one-electron energy gap inevitably affects magnitude of the $X_1$ exciton excitation energy, it will not change the conclusion that the $X_1$ exciton spontaneously forms and causes the excitonic instability. In more detail, one can see that the exciton excitation/binding energy first decreases/increases linearly, reaches its minimum/maximum near the experimental gap, and then reverses the variation trend, increasing/decreasing in complexity. Such a non-monotonic behavior reflects a sophisticated relationship between exciton energy structure and one-electron energy gap.

Akin to a superconductive state, the excitonic insulator state exists only at relatively low temperature. Going to high temperature, the many-body state becomes unstable more and more. Eventually, the exciton is dissociating at elevated temperature and the system presents the electronic properties characterized by its one-electron band structure. For example, a semi-hydrogenated graphene has an exciton binding energy exceeding its one-electron gap by $\sim$0.1 eV, while a transition temperature around 11 K is derived through an effective Hamiltonian calculation\cite{usgraphone}. It is well below the room temperature albeit high enough for a Bose condensate. Likewise, a similar performance can be expected for monolayer AlSb. In this regard, two-dimensional AlSb is more likely to present the one-electron properties instead of a spontaneous exciton condensate at room temperature. Owing to its comparable gap to the Si, it is natural to consider monolayer AlSb for applications in the next generation of electronic and optoelectronic devices.

\begin{table}
	\caption{Effective mass ($m^*$ in terms of free electron mass m$_0$), deformation-potential constant ($E_1$), elastic modulus ($C$), and carrier mobility ($\mu$) of monolayer AlSb at 298 K. All values are calculated on basis of the PBE band structure with the SOC included, as plotted in Fig. 2(b).}
	\begin{tabular}{cccccccccc}
		\hline
		carrier &  $m^*$ & $C$ (N/m) & $E_1$ (eV) & $\mu$ (cm$^2$/V$\cdot$s)  \\
		\hline
		electron & 0.07 & 62.95 & -10.38 & 1706 \\
		hole & 0.17 & 62.95 & -4.31 & 1695 \\
		\hline
	\end{tabular}
\end{table}

For this purpose, carrier mobility is a central quantity. Generally speaking, the carrier mobility needs to be at least on the level $\sim$500 cm$^2$/V$\cdot$s, in order to compete with commercially available silicon-based devices\cite{Ponce}. For example, the lower mobility limits monolayer MoS$_2$ transistor for high-performance applications\cite{Yoon}. Herein we consider the acoustic phonon scattering limited mobility within the deformation potential theory, which can be regarded as an upper limit for the experimentally achievable measurement. Under this method, the mobility $\mu$ is calculated by the formula\cite{Yoon,Kaasbjerg,Cai,Qiao,Dai}
\begin{equation}\label{(1)}
\mu = \frac{2e\hbar^3C}{3k_BT|m^*|^2E_1^2},
\end{equation}
where $T$ is the temperature and $m^*$ is the carrier effective mass. $E_1$ is the deformation potential constant and is obtained from the strain induced shift of conduction/valence-band extrema. $C$ is the elastic modulus, defined as $C$=$\partial^2E/\partial\delta^2$/$S^0$, where $E$ is the total energy, $\delta$ is the applied strain, and $S^0$ is the area of adopted supercell. To balance the computational efficiency and accuracy, here the mobility is estimated based on the PBE band including the SOC.

%fig05
\begin{figure}[tbp]
	\includegraphics[width=0.8\columnwidth]{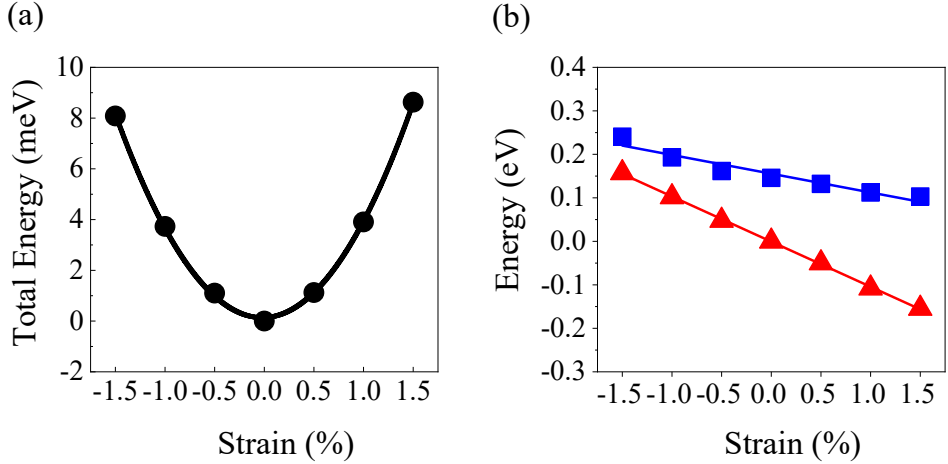}
	\caption{\label{fig:fig6} (Color online) (a) Change of total energy and (b) shift of band extrema, as a function of the isotropic in-plane strain. The total energy, as well as the valence band maximum, for the equilibrium case is set to zero for reference in the plots. In (b), red and blue lines correspond to the conduction band minimum and the valence band maximum, respectively. Note that herein the former is below the latter due to the band gap underestimation by the PBE [see Fig. 2(b)].}
\end{figure}

We first derive $m^*$ from the band shown in Fig. 2(b). They are 0.07 m$_0$ and 0.17 m$_0$, respectively for electrons and holes (m$_0$ denotes the free electron mass). We have examined the directional dependence of $m^*$ and an almost completely isotropic behavior is found. It is worth mentioning that the AlSb has both very light $m^*$ for electrons and holes, which is distinguished from other typical two-dimensional monolayer materials such as the MoS$_2$\cite{Yoon,Kaasbjerg,Cai}, phosphorene\cite{Qiao} and TiS$_3$\cite{Dai}.

In order to compute $E_1$ and $C$, strains up to 1.5\% are applied to the AlSb lattice and then the change of total energy, as well as the energy shift of band extrema, is measured relative to the strain. The results are shown in Fig. 5 and the obtained $E_1$ and $C$ are summarized in Table I. Inserting $m^*$, $E_1$ and $C$ into the formula (1), $\mu$ is computed to be around $\sim$1700 cm$^2$/V$\cdot$s, either for electrons or holes at 298 K (see Table I). In spite of its relatively lighter carrier $m^*$, the intrinsic mobility of monolayer AlSb is not high in comparison with other two-dimensional materials. It is only superior to the MoS$_2$\cite{Yoon,Kaasbjerg,Cai} and lower than those of phosphorene\cite{Qiao} and TiS$_3$\cite{Dai}. Anyway, such mobilities, together with its direct-gap feature and light $m^*$, render monolayer AlSb useful in applications of electronics and optoelectronics.

\vspace{0.3cm}
\textbf{IV. Conclusions}
\vspace{0.3cm}

In summary, we investigate the ground-state properties of newly synthesized two-dimensional monolayer III-V semiconductor AlSb by using the first-principles calculations in combination with Bethe-Salpeter equation. We find two attractive advantages of this material. On the one hand, an excitonic instability is revealed and thus it serves as a promising candidate for the long-sought excitonic insulator. On the other hand, monolayer AlSb can be used for high-performance electronic applications due to its light carrier effective mass and sufficient carrier mobility. We expect that our work could attract more attention to the two-dimensional form of traditional III-V semiconductors.

\vspace{0.3cm}
\textbf{Acknowledgments}
%\vspace{-1.5em}
\vspace{0.3cm}

This work was supported by the Ministry of Science and Technology of China (Grant No. 2020YFA0308800), the National Natural Science Foundation of China (Grant Nos. 12074034 and 11674071), and the Beijing Institute of Technology Research Fund Program for Young Scholars.


\begin{thebibliography}{90}%
\makeatletter
\bibitem{WangG} G. Wang, A. Chernikov, M. M. Glazov, T. F. Heinz, X. Marie, T. Amand, and B. Urbaszek, Rev. Mod. Phys. \textbf{90}, 021001 (2018).

\bibitem{Butler} S. Z. Butler, S. M. Hollen, L. Cao, Y. Cui, J. A. Gupta, H. R. Guti\'{e}rrez, T. F. Heinz, S. S. Hong, J. Huang, A. F. Ismach, E. Johnston-Halperin, M. Kuno, V. V. Plashnitsa, R. D. Robinson, R. S. Ruoff, S. Salahuddin, J. Shan, L. Shi, M. G. Spencer, M. Terrones, W. Windl, and J. E. Goldberger, ACS Nano \textbf{7}, 2898 (2013).

\bibitem{Bernardi} M. Bernardi, M. Palummo, and J. C. Grossman, Nano Lett. \textbf{13}, 3664 (2013).

\bibitem{Mott} N. F. Mott, Phil. Mag. \textbf{6}, 287 (1961).

\bibitem{Knox} R. X. Knox, in \emph{Solid State Physics}, edited by F. Seitz and D. Turnbull (Academic Press Inc., New York, 1963), Suppl. 5, p. 100.

\bibitem{Kohn} D. J\'{e}rome, T. M. Rice, and W. Kohn, Phys. Rev. \textbf{158,} 462 (1967).

\bibitem{Halperin} B. I. Halperin and T. M. Rice, Rev. Mod. Phys. \textbf{40,} 755 (1968).

\bibitem{KohnRMP} W.Kohn and D. Sherrington, Rev. Mod. Phys.\textbf{42}, 1 (1970).

\bibitem{Kogar} A. Kogar, S. Vig, M. S. Rak, A. A. Husain, F. Flicker, Y. I. Joe, L. Venema, G. J. MacDougall, T. C. Chiang, E. Fradkin, J. van Wezel, and P. Abbamonte, Science \textbf{358,} 1314 (2017).

\bibitem{usPRL} Z. Y. Jiang, Z. R. Liu, Y. C. Li, and W. H. Duan, Phys. Rev. Lett. \textbf{118}, 266401 (2017).

\bibitem{usEI}  Z. Y. Jiang, Y. C. Li, S. B. Zhang, and W. H. Duan, Phys. Rev. B \textbf{98}, 081408(R) (2018).

\bibitem{usHEI}  Z. Y. Jiang, Y. C. Li, W. H. Duan, and S. B. Zhang, Phys. Rev. Lett. \textbf{122}, 236402 (2019).

\bibitem{usgraphone} Z. Y. Jiang, W. K. Lou, Y. Liu, Y. C. Li, H. F. Song, K. Chang, W. H. Duan, and S. B. Zhang, Phys. Rev. Lett. \textbf{124}, 166401 (2020).

\bibitem{usDong} S. Dong and Y. C. Li, Phys. Rev. B \textbf{102}, 155119 (2020) .

\bibitem{Varsano} D. Varsano, M. Palummo, E. Molinari, and M. Rontani, Nat. Nanotechnol. 15, 367 (2020).

\bibitem{LiZ} Z. Li, M. Nadeem, Z. J. Yue, D. Cortie, M. Fuhrer, and X. L. Wang, Nano Lett. \textbf{19}, 4960 (2019).

\bibitem{Yan} R. Yan, A. Ourmazd, and K. F. Lee, IEEE Trans. Electron Devices \textbf{39}, 1704 (1992).

\bibitem{Liao} L. Liao, Y. -C. Lin, M. Bao, R. Cheng, J. Bai, Y. Liu, Y. Qu, K. L. Wang, Y. Huang, and X. Duan, Nature (London) \textbf{467}, 305 (2010).

\bibitem{Yoon} Y. Yoon, K. Ganapathi, and S. Salahuddin, Nano Lett.\textbf{11}, 3768 (2011).

\bibitem{Kaasbjerg} K. Kaasbjerg, K. S. Thygesen, and K. W. Jacobsen, Phys. Rev. B \textbf{85}, 115317 (2012).

\bibitem{Cai} Y. Q. Cai, G. Zhang, and Y. W. Zhang, J. Am. Chem. Soc.\textbf{136}, 6269 (2014).

\bibitem{Liu} H. Liu, A. T. Neal, Z. Zhu, Z. Luo, X. F. Xu, T. Tom\'{a}nek, and P. D. Ye, ACS Nano \textbf{8}, 4033 (2014).

\bibitem{Qiao} J. S. Qiao, X. H. Kong, Z. X. Hu, F. Yang, and W. Ji, Nat. Commun. \textbf{5}, 4475 (2014).

\bibitem{Balushi} Z. Y. Al Balushi, K. Wang, R. K. Ghosh, R. A. Vil\'{a}, S. M. Eichfeld, J. D. Caldwell, X. Y. Qin, Y.-C. Lin, P. A. DeSario, G. Stone, S. Subramanian, D. F. Paul, R. M. Wallace, S. Datta, J. M. Redwing, and J. A. Robinson, Nat. Mater. \textbf{15}, 1166 (2016).

\bibitem{Lucking} M. C. Lucking, W. Y. Xie, D.-H. Choe, D. West, T.-M. Lu, and S. B. Zhang, Phys. Rev. Lett. \textbf{120}, 086101 (2018).

\bibitem{Qin} L. Qin, Z. H. Zhang, Z. Y. Jiang, K. Fan, W. H. Zhang, Q. Y. Tang, H. N. Xia, F. Q. Meng, Q. H. Zhang, L. Gu, D. West, S. B. Zhang, and Y. S. Fu, https://arxiv.org/abs/2101.07431.

\bibitem{Yee} J. H. Yee, S. P. Swierkowski, and J. W. Sherohman, IEEE Trans. Nucl. Sci. \textbf{24}, 1962 (1977).

\bibitem{QE} P. Giannozzi, S. Baroni, N. Bonini, M. Calandra, R. Car, C. Cavazzoni, D. Ceresoli, G. L. Chiarotti, M. Cococcioni, I. Dabo, A. D. Corso, S. de Gironcoli, S. Fabris, G. Fratesi, R. Gebauer, U. Gerstmann, C. Gougoussis, A. Kokalj, M. Lazzeri, L. Martin-Samos, N. Marzari, F. Mauri, R. Mazzarello, S. Paolini, A. Pasquarello, L. Paulatto, C. Sbraccia, S. Scandolo, G. Sclauzero, A. P. Seitsonen, A. Smogunov, P. Umari, and R. M. Wentzcovitch, J. Phys.: Condens. Matter \textbf{21}, 395502 (2009).

\bibitem{PBE} J. P. Perdew, K. Burke, and M. Ernzerhof, Phys. Rev. Lett. \textbf{77,} 3865 (1996).

\bibitem{Hamann} D. R. Hamann, Phys. Rev. B \textbf{88,} 085117 (2013).

\bibitem{yambo} A. Marini, C. Hogan, M. Gr\"{u}ning, and D. Varsano, Comput. Phys. Commun. \textbf{180}, 1391 (2009).

\bibitem{Sahin} H. Sahin, S. Cahangirov, M. Topsakal, E. Bekaroglu,  E. Akturk, R. T. Senger, and S. Ciraci, Phys. Rev. B \textbf{80}, 155453 (2009).

\bibitem{Lei} S. D. Lei, L. H. Ge, S. Najmaei, A. George, R. Kappera, J. Lou, M. Chhowalla, H. Yamaguchi, G. Gupta, R. Vajtai, A. D. Mohite, and P. M. Ajayan, ACS Nano \textbf{8}, 1263 (2014).

\bibitem{Lucero} M. J. Lucero, T. M. Henderson, and G. E. Scuseria, J. Phys. Condens Matter. \textbf{24}, 145504 (2012).

\bibitem{Alibert} C. Alibert, A. Joulli\'{e}, A. M. Joulli\'{e}, and C. Ance, Phys. Rev. B \textbf{27}, 4946 (1983).

\bibitem{Ali} M. A. Ali, H. Aleem, B. Sarwar, and G. Murtaza, Indian J. Phys. \textbf{94}, 477 (2020).

\bibitem{Zhang} Y. Zhang, K. He, C. Z. Chang, C. L. Song, L. L. Wang, X. Chen, J. F. Jia, Z. Fang, X. Dai, W. Y. Shan, S. Q. Shen, Q. Niu, X. L. Qi, S. C. Zhang, X. C. Ma, and Q. K. Xue, Nat. Phys. \textbf{6}, 584 (2010).

\bibitem{Gonzalez} J. M. Gonzalez and I. I. Oleynik, Phys. Rev. B \textbf{94}, 125443 (2016).

\bibitem{Zhao} Z. Y. Zhao and Q. L. Liu, Catal. Sci. Technol. \textbf{8}, 1867 (2018).

\bibitem{Beian} M. Beian, M. Alloing, R. Anankine, E. Cambril, C. G. Carbonell, A. Lemaitre, and F. Dubin, Europhys. Lett. \textbf{119}, 37004 (2017).

\bibitem{Sangalli}  D. Sangalli, A. Ferretti, H. Miranda, C. Attaccalite, I. Marri, E. Cannuccia, P. Melo, M. Marsili, F. Paleari, A. Marrazzo, G. Prandini, P. Bonfa, M.O. Atambo, F. Affinito, M. Palummo, A. Molina-Sanchez, C. Hogan, M. Gruning, D. Varsano, and A. Marini, J. Phys.: Condens. Matter \textbf{31,} 325902 (2019).

\bibitem{Ponce} S. Ponc\'{e},  W. B. Li, S. Reichardt and F. Giustino. Rep. Prog. Phys. \textbf{83}, 036501 (2020).

\bibitem{Dai} J. Dai and X. C. Zeng, Angew. Chem. Int. Ed. \textbf{54}, 7572 (2015).

\end{thebibliography}
\end{document}